\documentstyle[12pt]{article}
\advance\voffset by 28 pt
\begin{document}
\thispagestyle{empty}
\noindent
{\flushright CERN-TH.6913/93\\}
\vspace{1cm}
\begin{center}
{\Large \bf Non-perturbative unification\\
in the light of LEP results}\\
\vspace{4mm}
{Biswajoy Brahmachari, Utpal Sarkar}\\
\vspace{2mm}
{\em Theory Group, Physical Research Laboratory,\\
Ahmedabad 380 009, India}\\
\vspace{4mm}
{and }\\
\vspace{4mm}
{K. Sridhar$^{*}$}\\
\vspace{2mm}
{\em Theory Division, CERN, CH-1211, Geneva 23, Switzerland.}\\
\end{center}

\vspace{.5cm}
\begin{abstract}
We consider an alternative to conventional grand unified theories
originally proposed by Maiani, Parisi and Petronzio, where owing to
the existence of extra fermion generations at some intermediate
scale, the gauge couplings become large at high energies. We
first comment on how the non-supersymmetric version of this
scenario is ruled out; we then
consider the two-loop evolution of couplings in the supersymmetric
extension of this scenario, and check whether such a scenario is
feasible in the light of the precise values of couplings now
available from LEP.
\end{abstract}

\vspace{2cm}
\noindent
\vspace{1cm} $^{*)} $ sridhar@vxcern.cern.ch\\
CERN-TH.6913/93\\
June 1993\\

\vfill
\clearpage
\setcounter{page}{1}
\pagestyle{plain}
Grand unified theories (GUTs) offer the possibility of unifying the
SU(3), SU(2) and U(1) gauge groups of the standard model
into one large group at a high energy scale, $M_U$. This scale
is determined as the intersection point of the SU(3), SU(2)
and U(1) couplings. The particle content of the theory
completely determines the variation of the couplings with energy.
Given the particle content of the theory, therefore, one can
evolve the couplings determined at low energies to determine whether
there is unification.

The determination of the couplings at LEP has important consequences
for grand unified theories. The precise determination of the
Weinberg angle, $\theta_W$, and the strong coupling, $\alpha_s$,
at the $Z$-peak has helped in putting rather stringent constraints
on unified models \cite{amaldi,lepgut}. In particular, it has been
found \cite{amaldi} that in the standard model with three fermion
generations and one Higgs doublet, the couplings do not meet at a
single point at high energies. In contrast, in the minimal
supersymmetric extension of the standard model (with three
generations and two Higgs doublets), a single intersection point
obtains at about $10^{16}$~GeV.
The compatibility of this simple supersymmetric GUT with the
couplings determined from LEP is remarkable. Nonetheless, it is
important to study other models, which are alternatives to grand
unification, and see whether they are viable in the light of the
available experimental information on couplings.

An interesting alternative to GUTs was proposed by Maiani, Parisi
and Petronzio \cite{mpp} several years ago. In this scheme, the
couplings enter a non-perturbative phase at a high energy scale,
i.e. the theory is asymptotically divergent. Starting from the
renormalisation group equation for a coupling $\alpha$,
\begin{equation}
\label{e1}
{d\alpha \over dt} = \beta (\alpha ),
\end{equation}
where $\beta(\alpha)$ is the beta function and $t=\mbox{\rm ln}
(Q^2/\mu^2)$, $\mu$ being some reference scale, we obtain
\begin{equation}
\label{e2}
t = \int_{\alpha(\mu)}^{\alpha(Q^2)} {d\alpha \over \beta(\alpha)}.
\end{equation}
For $\beta(\alpha) > 0$ (asymptotically divergent theory) there is
a value of $t$, given by
\begin{equation}
\label{e3}
t = \int_{\alpha(\mu)}^{\infty} {d\alpha \over \beta(\alpha)}
< \infty ,
\end{equation}
for which $\alpha \rightarrow \infty$. If perturbation theory is to
be valid at all energy scales, we require $\alpha(\mu)=0$, so
that $t_c=\infty$, $\alpha(\mu)=0$ is the infra-red fixed
point. But if $\alpha(\mu)\ne 0$ but small, i.e. it is sufficiently
close to the infra-red fixed point, then there is a finite cut-off in
energy beyond which the theory is non-perturbative.

In Ref.~\cite{mpp}, it was assumed that the standard
SU(3)$\times$SU(2)$\times$U(1) theory, due to new fermion generations
that get switched on around the weak scale $\Lambda_F = 250$~GeV,
is asymptotically
divergent beyond $\Lambda_F$. The couplings $\alpha_{1,2,3}$ are
sufficiently close to zero at $\Lambda_F$ but not quite zero. As
a consequence, the theory is cut off at a scale $\Lambda$.
At this scale, the most interesting situation is that not just
one but all three couplings are large, i.e. of O(1). In fact, it
has been shown \cite{hung} that such a non-perturbative scenario
exhibits a "trapping" mechanism, whereby if one of the couplings
grows large, the other couplings will also increase. This effect,
by means of which all three couplings are large and of the
same order of magnitude at $\Lambda$, leads to what is called
non-perturbative unification. In Ref.~\cite{mpp} the cut-off
scale $\Lambda$ was assumed to be the Planck scale; however,
in subsequent studies \cite{cabfar,grunberg}, $\Lambda$ was
determined to be of the order of $10^{15}$~--$10^{17}$~GeV. Since
the low-energy couplings are close to the infra-red fixed point,
they are insensitive to the values of the couplings at the scale
$\Lambda$.

One natural extension of the above scenario is the inclusion of
supersymmetry. This was first considered in Ref.~\cite{cabfar},
and was later discussed in Refs.~\cite{grunberg,mp}. Other than
solving the hierarchy problem, the inclusion of supersymmetry is
attractive because it provides a framework for the existence
of new particles needed to make the theory asymptotically
divergent. In the case of the simplest $N=1$ supersymmetric
extension of the scenario, it suffices to consider $n_f=5$, where
$n_f$ is the number of fermion generations.

In this letter, we use the recent LEP values to check whether any
strong constraints on the non-perturbative unification scenario
can be obtained.
The values of $\mbox{\rm sin}^2\theta_W$ and $\alpha_s$ from LEP
are very
precise compared to that available from older experiments. One
strong constraint is on the number of extra chiral generations.
The present limit on the oblique parameters S, T and U allows only
three chiral fermion generations, while the vectorial
generations are not constrained. Thus in addition to the three
chiral fermion generations we are allowed to have only an even
number of generations.

We shall first specify the supersymmetric non-perturbative unification
scenario in detail. While discussing the results we shall also
comment on the results of the non-supersymmetric case. We
consider an SU(3)$\times$SU(2)$\times$U(1)
supersymmetric gauge theory with the assumption that an $N=1$
supersymmetry holds above the scale $\Lambda_s$. We assume $n_f=5$
supersymmetric generations and two Higgs supermultiplets. In the
discussion of the non-supersymmetric case we shall consider one
Higgs scalar and $n_f=8 \;\;\mbox{\rm and}\:\: 9$. From
the requirement that the Yukawa couplings do not become arbitrarily
large, a bound on the fermion masses can be obtained \cite{cmpp,bdm}.
This bound is that fermion masses are, in general, smaller than
200--250~GeV. We assume that the extra fermion generations, which are
required for the theory to be asymptotically divergent, are of the
order of 250 GeV in mass.

Having specified the theory we can now address the question of the
evolution of the three couplings. The two-loop renormalisation group
equations for the couplings are given by the following coupled
differential equations:
\begin{equation}
\label{e4}
\mu{d\alpha_i(\mu) \over d\mu} = {1 \over 2\pi} \biggl \lbrack
a_i + {b_{ij} \over 4\pi}\alpha_j(\mu)+{b_{ik} \over
4\pi} \alpha_k(\mu) \biggr \rbrack \alpha_i^2(\mu) +
{2b_{ij} \over (4\pi)^2} \alpha_i^3(\mu) ,
\end{equation}
where $i,\ j,\ k = 1,\ 2,\ 3$ and $i \ne j \ne k$, and
$a_i$ and $b_{ij}$ are the one- and two-loop beta function coefficients.
In the range of energies between $M_Z$ and the supersymmetric threshold,
$M_s$, we use the non-supersymmetric beta functions to evolve the
couplings, whereas from $M_s$ onward the supersymmetric beta functions
are effective. We retrieve the result for the non-supersymmetric
scenario by taking $M_s = \Lambda_{MPP}$ and large $n_f$.

In the non-supersymmetric case the one-loop beta function
coefficients are \cite{jones}
\begin{eqnarray}
\label{e5}
b_j&=&\pmatrix{0 \cr -{22\over3} \cr -11\cr}+n_f\pmatrix{{20\over9} \cr
{4\over3} \cr {4\over3}\cr}+n_h\pmatrix{{1\over6} \cr
{1\over6} \cr {0}\cr}
\end{eqnarray}
while the two-loop beta functions are
\begin{eqnarray}
\label{e6}
a_{ij}&=&-\pmatrix{0&0&0 \cr0& {136\over3}&0 \cr0&0&
102\cr}+n_f\pmatrix{{95\over27}&1&{44\over9} \cr
{1\over3}&{49\over3}&4 \cr
{11\over18}&{3\over2}&{76\over3}\cr}+n_h\pmatrix{{1\over2}&{13\over6}&0
\cr  {1\over2}&{13\over6}&0 \cr 0&0&0\cr}
\end{eqnarray}
In the supersymmetric case the one-loop beta functions take the
form \cite{jones}
\begin{eqnarray}
\label{e7}
b_j&=&\pmatrix{0 \cr -6 \cr -9\cr}+n_f\pmatrix{{10\over3} \cr
2 \cr 2\cr}+n_h\pmatrix{{1\over2} \cr
{1\over2} \cr {0}\cr}
\end{eqnarray}
while the two-loop beta functions are
\begin{eqnarray}
\label{e8}
a_{ij}&=&-\pmatrix{0&0&0 \cr0& 24&0 \cr0&0&
54\cr}+n_f\pmatrix{{190\over27}&2&{88\over9} \cr
{2\over3}&14&8 \cr
{11\over9}&3&{68\over3}\cr}+n_h\pmatrix{{1\over2}&{3\over2}&0
\cr  {1\over2}&{7\over2}&0 \cr 0&0&0\cr}
\end{eqnarray}
In all these equations, $n_f$ and $n_h$ denote the number of
fermion generations and the number of Higgs doublets
respectively.

We integrate the coupled differential equations in Eq.~(\ref{e4})
numerically, with the initial values of the three couplings
$\alpha_{1,2,3}$ taken to be of O(1) at the unification scale
$\Lambda$. What we do in practice is to evolve downwards using
the renormalisation group equations for several values of $\Lambda$,
and check what the predicted values of the couplings at the scale
$M_Z$ are. The extra fermion generations are assumed to contribute
to the beta functions for all energies greater than 250~GeV.

We shall first comment on the non-supersymmetric scenario and
then present our main result, namely the supersymmetric
extension. In this case we find that for $n_f \leq 8$,
$\alpha_2(M_z)$ remains too small, and that $\alpha_{1,2}(M_z)$ falls
within the experimental bound for $n_f \geq 9$. But for $n_f
\geq 9$ the strong coupling constant evolves extremely fast and
$\alpha_3(M_z)$ becomes too
large. Thus the precision LEP data rule out the
non-supersymmetric scenario completely.

The results of the computation for the supersymmetric version
are shown in Fig.~1, where
$\alpha_{1,2,3}(M_Z)$ are shown as a function of $\Lambda$. The
solid, dashed and dotted curves are for $M_s=250$~GeV, 1.2~TeV and
5~TeV, respectively. The horizontal lines in the figures show the
upper and lower bounds on the couplings at $M_Z$ as determined
by the LEP experiment. These are as follows \cite{lep}:
\begin{eqnarray}
\label{e9}
\alpha_1 &=& 0.0101322 \pm 0.000024 \nonumber\\
\alpha_2 &=& 0.03322 \pm 0.00025 \nonumber\\
\alpha_3 &=& 0.120 \pm 0.006.
\end{eqnarray}
It is clear from the figure that the
non-perturbative unification scheme is certainly viable if we have
$M_s=1.2$~TeV and $\Lambda$ close to $0.78 \times 10^{17}$~GeV. We
have checked that the range of values allowed is $M_s = 1.2 \pm
0.2$~TeV and $\Lambda=$(0.7--0.8)$\times 10^{17}$~GeV. We have
also checked that the couplings at $M_Z$ are not sensitive to
the choice of the couplings at $\Lambda$. We have checked this
by varying these from 0.75 to 10.

Let us now summarise our results. We have studied the
non-perturbative unification scenario first proposed
by Maiani, Parisi and Petronzio. We point out that the
non-supersymmetric version of this scenario is ruled out by LEP
data. However, the supersymmetric extension of this scenario remains
a viable alternative to conventional grand unified theories and is
capable of predicting the precision values of couplings determined
from LEP. Our numerical results show that the non-perturbative
scale, $\Lambda$, at which all couplings are large, is around
0.7--0.8$\times 10^{17}$~GeV, with the supersymmetric threshold
$M_s$ around 1.0--1.4~TeV. If the scale $M_s$ gets either larger
or smaller it is then not possible to reproduce the values of the
couplings at $M_Z$. We should note that the agreement with the data
is obtained only for a constrained range of parameters of this
scenario. In principle, the effect of higher-order corrections
could be large and this may ruin the agreement. It is also likely
that more accurate measurements of the strong coupling $\alpha_3$
at low energies may be sufficient to either put strong constraints
or completely rule out this scenario. It is nevertheless interesting
that this scenario, at the two-loop level, is a possible alternative
to conventional grand unification.

\newpage
\section*{Figure caption}
\renewcommand{\labelenumi}{Fig. \arabic{enumi}}
\begin{enumerate}
\item   
The couplings
$\alpha_{1,2,3} (M_Z)$ as a function of the non-perturbative unification
scale $\Lambda$. The solid, dashed and dotted curves are for
$M_s$=250~GeV, 1.2~TeV and 5~TeV, respectively. The horizontal lines
in the figures show the experimentally allowed upper and lower bounds
on the couplings at $M_Z$.
\end{enumerate}
\end{document}